\documentclass{vgtc}                          

\graphicspath{{figures/}{pictures/}{images/}{./}} 

\usepackage{times}                     

\usepackage{tabu}                      
\usepackage{booktabs}                  
\usepackage{lipsum}                    
\usepackage{mwe}                       
\usepackage{multirow}

\usepackage{mathptmx}                  

\newcommand{\colour}[1]{\textcolor{black}{#1}}  

\onlineid{0}

\vgtccategory{Research}

\vgtcinsertpkg

\title{How Long Does It Take to Alleviate Discomfort? A Preliminary Study on Reducing Cybersickness in Novice Users}

\author{
  Zhengxin Zhang\thanks{School of Design, Inner Mongolia Normal University, Hohhot, China. E-mail: robinzhang2015@iCloud.com}
  \and
  Shufang Qian\thanks{School of Design, Inner Mongolia Normal University, Hohhot, China. E-mail: qianshufang@imnu.edu.cn}
  \and
  Yi Wang\thanks{School of Information Technology, Deakin University, Melbourne, Australia. E-mail: xve@deakin.edu.au}
  \and
  Xiao Liu\thanks{School of Information Technology, Deakin University, Melbourne, Australia. E-mail: xiao.liu@deakin.edu.au}
  \and
    Chetan Arora\thanks{Faculty of Information Technology, Monash University, Melbourne, Australia. E-mail: chetan.arora@monash.edu}
  \and
  Thuong Hoang\thanks{School of Information Technology, Deakin University, Melbourne, Australia. E-mail: thuong.hoang@deakin.edu.au}
  \and
  Jingjing Zhang\thanks{School of Design, Inner Mongolia Normal University, Hohhot, China. E-mail: zjj18606533706@163.com}
  \and
  Henry Been‑Lirn Duh\thanks{School of Design, The Hong Kong Polytechnic University, Hong Kong, China. E-mail: henry.duh@polyu.edu.hk}
}

\abstract{

Cybersickness significantly impacts the user experience in VR applications. Locomotion tunneling is a widely adopted technique for mitigating cybersickness in susceptible users. However, there is a lack of research investigating the effects of prolonged use of locomotion tunneling on cybersickness symptoms among novice users. To fill this gap, we used VRChat as our experimental platform. We recruited 24 novice VR users, defined as participants with no prior experience using immersive virtual environments. We collected five days of data within a one-week period. The results indicated that participants exhibited significant mitigation to cybersickness by Day 4. However, a change in the VR scene on Day 5 led to a notable increase in cybersickness symptoms. Qualitative feedback revealed participant-perceived causes of cybersickness and suggested that the effectiveness of locomotion tunneling was limited in some scenarios. Finally, we discussed the limitations of the study and proposed directions for future research.

} 


\begin{document}



\maketitle 

\section{Introduction}

Virtual Reality (VR) is rapidly expanding across domains such as gaming, healthcare, and education, offering users unprecedented immersive interactions \cite{WangYi2025}. However, cybersickness remains a significant barrier to the widespread adoption of VR technology. Cybersickness occurs when there is a misalignment between visual and vestibular cues and the body's actual or expected motion \cite{zhang2016motion}. This sensory conflict triggers symptoms such as dizziness and nausea, especially in novice users \cite{zhang2016motion}. From a neurological standpoint, such mismatches are believed to challenge the brain’s ability to reconcile conflicting inputs from the visual and vestibular systems \cite{oman2014brainstem}. Although prior research on traditional screen-based media \cite{zhang2016motion}, such as 3D games or flight simulators, has shown that rapid scene movements can induce similar discomfort, fully immersive VR tends to exacerbate this mismatch between the visual and vestibular systems.

Existing solutions mainly focus on hardware, software design, and assistive technology, such as improving display resolution, reducing system latency, and adopting dynamic fields of view (FOV) and texture-blurring techniques \cite{luks2019investigating}. Although these methods have proven effective in reducing the incidence and severity of cybersickness, they do not fully address the unique vulnerability of novice users during their initial exposure to VR technology. This heightened susceptibility often leads to discomfort and may discourage prolonged engagement or future use of immersive systems. While system-level technical approaches are valuable, they often overlook individual differences in symptom manifestation and adaptation processes among users \cite{kourtesis2024cybersickness}. Moreover, the duration and key factors of novice users’ adaptation to immersive environments remain unclear. Additionally, limited research has investigated the progression of cybersickness during prolonged exposure to existing VR applications.


To investigate cybersickness mitigation approaches among novice VR users, we conducted a five-day experiment with 24 undergraduate students with no prior experience in VR. Participants were categorized into mild and high cybersickness groups based on their Day 1 Simulator Sickness Questionnaire (SSQ) \cite{kennedy1993simulator} scores. Participants joined in VR Jetski Rush and VR C1 Runner within VRChat \cite{VRChat2024}, using tailored levels of locomotion tunneling, a visual mitigation technique based on individual susceptibility. Daily cybersickness levels were measured using the SSQ. Our objectives were to investigate the progression of cybersickness with repeated VR usage and to assess how specific mitigation strategies, such as locomotion tunneling, influence symptom reduction. These findings provide practical guidelines for improving VR accessibility and enhancing user retention, especially among first-time users. By analyzing both quantitative and qualitative data, we gained a deeper understanding of the mechanisms underlying cybersickness among novice VR users.

\section{Related Work}


Cybersickness arises from sensory conflict, particularly when visual inputs diverge from vestibular and proprioceptive signals \cite{oman2014brainstem}. This mismatch can lead to symptoms such as nausea, dizziness, and fatigue. It is especially prevalent in VR applications that involve rapid movements, sudden scene transitions, or artificially rendered environments, such as flight simulators, 3D games, and amusement rides. Such sensory conflicts are particularly pronounced in scenarios featuring artificial locomotion, which is common in racing or fast-paced games \cite{zhang2016motion}. To mitigate cybersickness, researchers have proposed techniques that reduce the mismatch between visual and inertial cues. These include visual stabilization (e.g., keeping the background static while the foreground moves) \cite{Nie2020, zhang2016motion}, image blurring, dynamic FOV restriction \cite{Wu2022AdaptiveFOVRestriction}, and gaze-contingent image deformation to reduce peripheral motion perception \cite{Groth2024GazeContingentDeformation}. Such strategies aim to harmonize sensory inputs and enhance user comfort in immersive environments. Meanwhile, locomotion tunneling is a technique that reduces users’ peripheral FOV during movement in VR, aiming to minimize visual flow and reduce sensory conflict. A ``high locomotion tunneling" condition applies a strong vignette with a narrow central view, while ``low locomotion tunneling'' offers a mild vignette, retaining more peripheral visibility. In contrast, the ``no locomotion tunneling'' condition presents the full FOV with no visual restrictions. While system-level  technical approaches can partially alleviate cybersickness, individual variability remains a critical factor in the adaptation process. Moreover, there has been limited exploration of how cybersickness evolves over time, particularly among novice VR users.

\section{\colour{Formative Study}}

\begin{figure*}
    \centering
    \includegraphics[width=\linewidth]{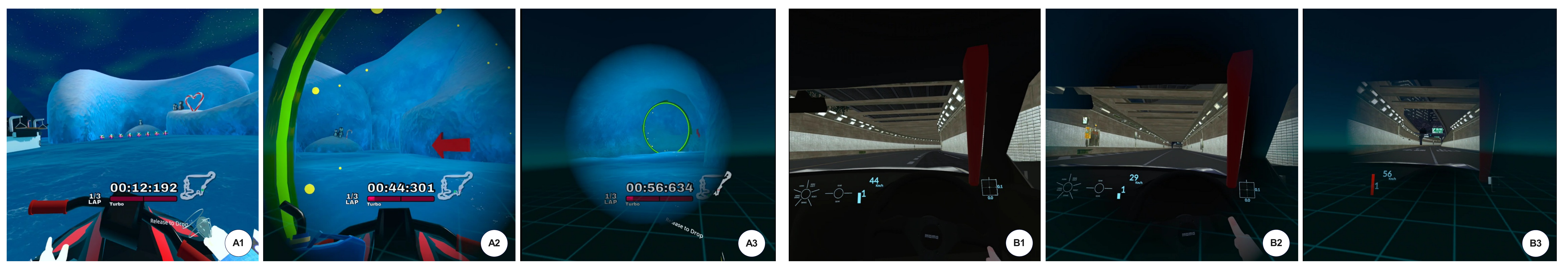}
    \caption{\colour{Two VR games (A-VR Jetski Rush, and B-VR C1 RUnner) of the different types of locomotion tunnelings. A1 and B1 are the original virtual environments (no locomotion tunneling). A2 and B2 are the low locomotion tunneling. A3 and B3 are the high locomotion tunneling. }}
    \label{fig:1}
\end{figure*}


We conducted a formative study to investigate whether participants with varying levels of cybersickness could alleviate their symptoms while engaging in VR games with faster movement speeds. Our preliminary analysis of VR accessibility reviews from the Meta and Steam stores revealed that many users who reported cybersickness identified themselves as novice VR users. Several reviews noted that their symptoms diminished over time through repeated use of applications such as VRChat, although similar applications continued to trigger discomfort. To explore this phenomenon, we recruited novice VR users to engage with an existing VR application (VRChat) over the course of four days, followed by a change in the VR scene on the fifth day. The aim of this experiment was to examine whether novice users could adapt to and overcome cybersickness under different conditions. The study was approved by the Research Ethics Committee at our university.

\subsection{Hypothesis}

\colour{Our study explored how these settings differentially affect cybersickness. Based on this, we propose the following three hypotheses:} 

\colour{\textbf{H1. }High locomotion tunneling will significantly reduce cybersickness symptoms (\textbf{H1.1}) and prevent symptom recurrence following VR scene transitions (\textbf{H1.2}).}

\colour{\textbf{H2. }Low locomotion tunneling will offer moderate reduction in cybersickness symptoms (\textbf{H2.1}) and partial prevention of symptom recurrence following VR scene transitions (\textbf{H2.2}).}

\colour{\textbf{H3. }No locomotion tunneling will result in the highest cybersickness symptoms (\textbf{H3.1}) and increased likelihood of symptom recurrence after scene transitions (\textbf{H3.2}).}

\subsection{Participants}


\colour{We initially recruited 27 undergraduate students (13 males and 14 females). All participants were novice users with no prior VR experience. Two participants did not report any cybersickness symptoms during the study and were excluded from the final analysis (the SSQ score is below 1). One additional participant (from the high cybersickness group) withdrew on the third day. These three participants were excluded as they did not meet the inclusion criteria. The final sample included 24 participants aged between 18 and 22 years (\textit{M} = 18.96, \textit{SD} = 1.10), with 13 assigned to the mild cybersickness group (6 males, 7 females) and 11 to the high cybersickness group (5 males, 6 females).} Participants with an average score between 1 and 5 were classified as experiencing mild cybersickness, while participants with an average score between 6 and 10 were classified as experiencing high cybersickness. Participants were divided into four groups, two of which served as control groups. Participants with mild susceptibility to cybersickness were assigned to either a low locomotion tunneling condition or a no locomotion tunneling condition. Conversely, those with high susceptibility were assigned to either a high locomotion tunneling condition or a no locomotion tunneling condition.



\subsection{VR Game Selection and Conditions}

\colour{We used a between-subject design for our experiment. We selected VRChat \cite{VRChat2024} as the VR experiment platform because it supports remote collaboration and communication among multiple users.} It also provided a wide range of virtual entertainment environments. Our goal was to select VR games that are likely to induce cybersickness, such as VR racing games and VR skiing games. Further, we selected VR C1 Runner and VR Jetski Rush. \colour{Based on our testing of the two VR games, both exhibited fast-paced visual motion and continuous navigation factors known to increase the risk of cybersickness due to intensified optic flow and sensory conflict \cite{ng2020study}. }We then confirmed that both VR games supported smooth locomotion, a joystick-based movement mechanism that simulates walking in VR without requiring physical displacement.  Moreover, locomotion tunneling was an accessibility feature that included three modes: low, high, and none. It reduced cybersickness for users while playing VR games. Figure \ref{fig:1} A2 \& B2 and A3 \& B3 show the settings for low and high levels, respectively. Figure \ref{fig:1} A1 and B1 show the original scenes. We deployed VRChat with VR games on two Meta Quest 3 headsets and two Meta Quest 3s headsets. 




\subsection{Procedure}

\textbf{Introduction and Training (Day 1).} We welcomed all participants and introduced the purpose of our study, data privacy, and potential ethical risks. We reminded all participants that they could stop or withdraw from the experiment at any time if they felt uncomfortable. In such cases, their data were considered invalid and excluded from the analysis. Our experiment lasted for five days, with no more than 30 minutes per day (excluding the completion of the questionnaire). Participants could schedule their preferred times with the researchers after completing the experiment. All participants must complete all tasks within seven days. We then introduced the basic operations of VRChat and the use of the Meta Quest headset. This stage did not exceed ten minutes. In each group, participants were required to complete two VR simulations: VR Jetski Rush and VRChat C1 Runner. Prior to each simulation, participants received a detailed explanation of the basic controls and objectives. They were then instructed to compete for the highest possible ranking. If a participant experienced severe discomfort at any point during a simulation, they were permitted to exit immediately and complete the corresponding questionnaire. Participants who did not report discomfort were permitted to continue each simulation for up to thirty minutes.Four researchers were responsible for organizing the experiment.



\textbf{Day 2 to Day 5. }From Day 2 to Day 4, all participants used VR Jetski Rush. On the final day, to preliminarily assess whether participants could alleviate cybersickness and whether the solution was applicable to other similar scenarios, we used VR C1 Runner on Day 5. 

\subsection{Measurement}

We used the SSQ (a 10-point Likert scale) \cite{kennedy1993simulator} to measure participants' levels of cybersickness. We also designed a qualitative questionnaire to gather honest feedback from participants, which included six open-ended questions. 



\section{Results}

\begin{figure}[t]
    \centering
    \includegraphics[width=\linewidth]{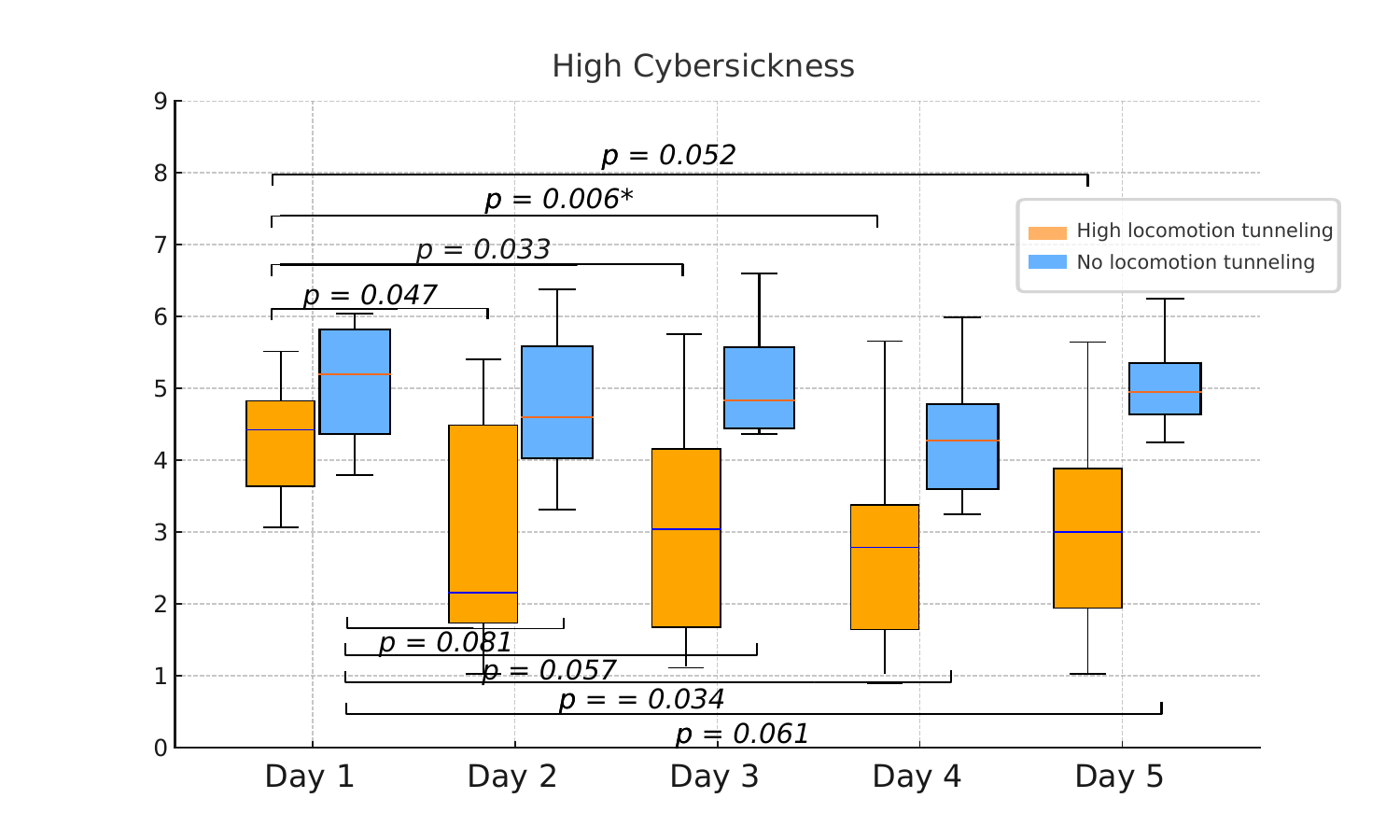}
    \caption{\colour{The results of high cybersickness from Day 1 to Day 5 in both ``high locomotion tunneling'' (orange) and ``no locomotion tunneling'' (blue) conditions. } }
    \label{fig:enter-label}
\end{figure}
\begin{figure}[t]
    \centering
    \includegraphics[width=\linewidth]
    {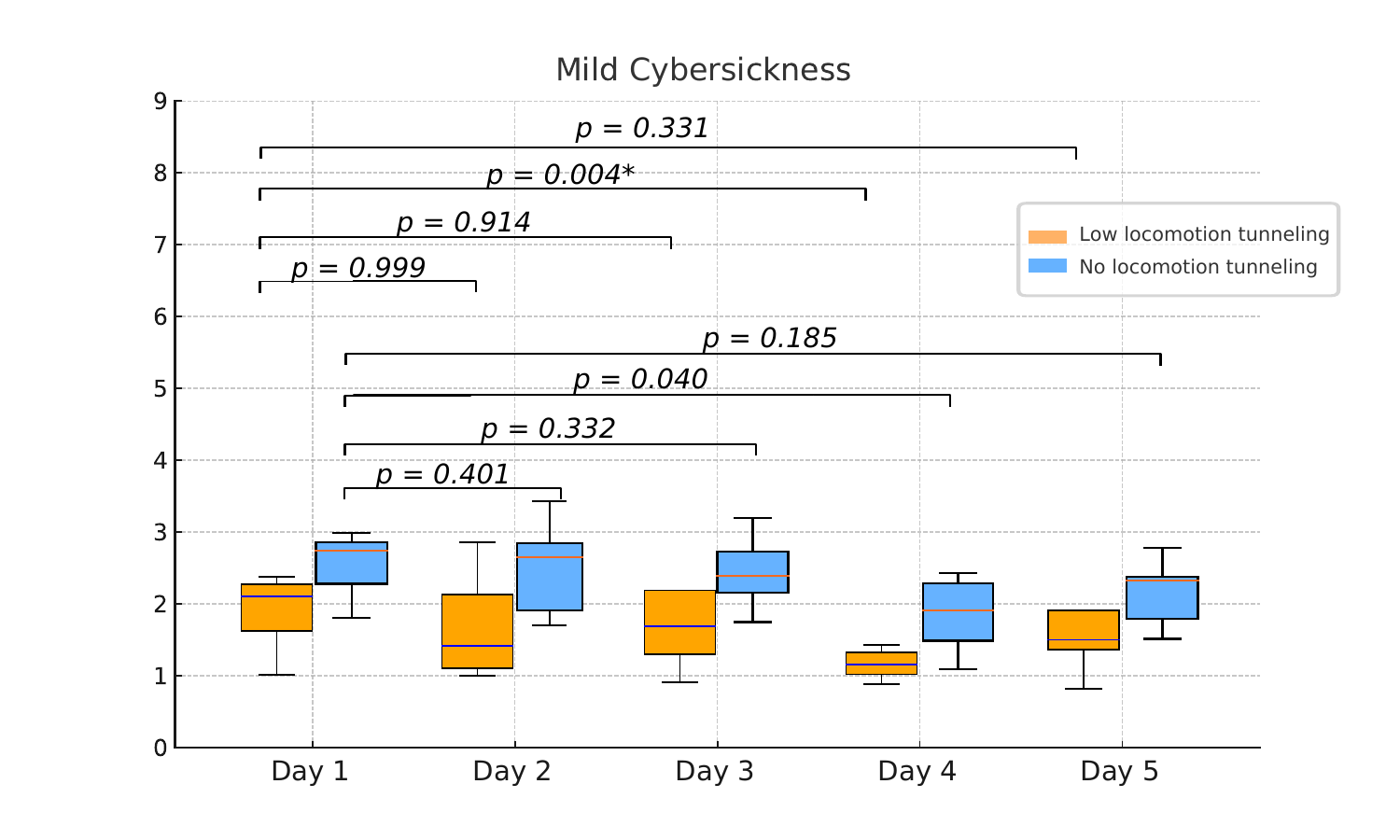}
    \caption{\colour{The results of mild cybersickness from Day 1 to Day 5 in both ``low locomotion tunneling'' (orange) and ``no locomotion tunneling'' (blue) conditions. }} 
    \label{fig:enter-label2}
\end{figure}

We employed the Wilcoxon signed-rank test to analyze experimental data collected over five days under different conditions. We used the results from the first day as a baseline for comparison with the results of the subsequent four days. Additionally, we used the Bonferroni correction to derive a \textit{$\alpha$} = 0.0125. Furthermore, we used affinity diagramming to organize the participants’ feedback. Figure \ref{fig:enter-label} and \ref{fig:enter-label2} show the results of the analysis.


\subsection{Day 1 vs. Other Four Days } 


\textbf{Day 1 vs. Day 2. } For the high cybersickness group, no significant difference was observed between Day 1 and Day 2  when using either the high locomotion tunneling condition (\textit{p} = .047) or the no locomotion tunneling condition (\textit{p} = .081). Descriptive statistics indicated a lower score on Day 2 when using a high locomotion tunneling condition (\textit{Mean} = 4.11, \textit{SD} = 1.08). For the mild cybersickness group, no significant difference was observed between Day 1 and Day 2 when using either the low locomotion tunneling condition (\textit{p} = .999) or the no locomotion tunneling condition (\textit{p} = .401). Descriptive statistics indicated a lower score on Day 2 when using the low locomotion tunneling condition (\textit{Mean} = 1.45, \textit{SD} = 0.60).

\textbf{Day 1 vs. Day 3. } For the high cybersickness group, no significant difference was observed between Day 1 and Day 3 when using either the high locomotion tunneling condition (\textit{p} = .033) or the no locomotion tunneling condition (\textit{p} = .057). Descriptive statistics indicated a lower score on Day 3 when using the high locomotion tunneling condition (\textit{Mean} = 3.10, \textit{SD} = 0.98). For the mild cybersickness group, no significant difference was observed between Day 1 and Day 3 when using either the low locomotion tunneling condition (\textit{p} = .914) or the no locomotion tunneling condition (\textit{p} = .332). Descriptive statistics indicated a lower score on Day 3 when using the low locomotion tunneling condition (\textit{Mean} = 1.52, \textit{SD} = 0.58 ).

\textbf{Day 1 vs. Day 4. }For the high cybersickness group, a significant difference was observed between Day 1 and Day 4 when using the high locomotion tunneling condition (\textit{p} = .006, \textit{Z} = 2.521). In contrast, no significant difference was observed when using the no locomotion tunneling condition (\textit{p} = .034). Descriptive statistics indicated a lower score on Day 4 when using the high locomotion tunneling condition (\textit{Mean} = 2.87, \textit{SD} = 0.90). For the mild cybersickness group, a significant difference was observed between Day 1 and Day 4 when using either the low locomotion tunneling condition (\textit{p} = .004, \textit{Z} = 2.651). In contrast, no significant difference was observed between Day 1 and Day 4 when using the no locomotion tunneling condition (\textit{p} = .040). Descriptive statistics indicated a lower score on Day 4 when using the low locomotion tunneling condition (\textit{Mean} = 1.20, \textit{SD} = 0.50).


\textbf{Day 1 vs. Day 5. }For the high cybersickness group, no significant difference was observed between Day 1 and Day 5 when using either the high locomotion tunneling condition (\textit{p} = .052) or the no locomotion tunneling condition (\textit{p} = .061). Descriptive statistics indicated a lower score on Day 5 when using the high locomotion tunneling condition (\textit{Mean} = 3.18, \textit{SD} = 0.80). For the mild cybersickness group, no significant difference was observed between Day 1 and Day 5 when using either the low locomotion tunneling condition (\textit{p} = .331) or the no locomotion tunneling condition (\textit{p} = .185). Descriptive statistics indicated a lower score on Day 5 when using the low locomotion tunneling condition (\textit{Mean} = 1.89, \textit{SD} = 0.42).


\subsection{Qualitative Feedback}

Several participants attributed their discomfort to the interaction method, particularly joystick-based smooth locomotion and smooth turning. They reported that these continuous movement mechanisms, especially when used during high-speed sequences, induce a sense of disorientation. Other participants identified sudden perspective shifts, such as those occurring when entering or exiting vehicles, as a source of discomfort. Environmental factors were also identified, including high light intensity (e.g., overly bright scenes), loud in-game audio, and an excessive number of visual elements, all of which were perceived as contributing to sensory overload. Although many participants considered that locomotion tunneling could alleviate certain symptoms, particularly those related to brightness and motion, it was not universally effective. Participants with high susceptibility to cybersickness often found high locomotion tunneling to be ineffective and visually disruptive. In contrast, some participants with low susceptibility reported that low locomotion tunneling had minimal impact, suggesting that repeated exposure and gradual adaptation may be more beneficial. Nonetheless, several participants noted that the slight peripheral dimming used in low locomotion tunneling did not negatively affect their experience.



\section{Discussion}


\subsection{Day 1 vs. Other Four Days}

\colour{Based on the quantitative data, participants demonstrated a gradual reduction in cybersickness symptoms over the five-day period, with a statistically significant decrease observed on Day 4. On that day, both the mild and high cybersickness groups exhibited noticeable symptom mitigation, supporting hypotheses H1.1 and H2.1. However, no significant improvement was observed on Days 2 or 3, indicating that the adaptation process was not linear. Consequently, H1.1 and H2.1 were not supported on Days 2 and 3, indicating that high locomotion tunneling did not reduce cybersickness symptoms, and low locomotion tunneling did not provide a moderate reduction in symptoms during these days.} Despite the lack of statistical significance, the downward trend in symptoms on those days suggests progressive adaptation.

\colour{Importantly, the use of locomotion tunneling yielded mixed effects across different types of user groups. While high locomotion tunneling initially reduced discomfort for some highly susceptible participants, others found it visually disruptive or ineffective, revealing a trade-off between symptom relief and immersion quality. In contrast, participants with mild susceptibility often reported that low locomotion tunneling had minimal impact, attributing symptom reduction more to repeated exposure and increased familiarity with the environment. These divergent responses suggest that locomotion tunneling is not a universally optimal solution, its effectiveness is influenced by the individual factors such as prior susceptibility and visual masking.}

On Day 5, when a different VR game was introduced, participants across all conditions experienced a resurgence of cybersickness symptoms. This result did not support hypotheses H1.2 and H2.2. Descriptive statistics indicated that both mild and high susceptibility participants were affected. Participant feedback suggested that the abrupt change in the virtual environment contributed to this increase, likely due to differences in scene fidelity, brightness, and the density of 3D objects. 

\colour{Participants with high cybersickness exhibited lower scores when no locomotion tunneling was used on Day 4, partially supporting hypothesis H3.1. However, symptoms intensified again for both groups on Day 5, and hypothesis H3.2 was not supported. While some participants appreciated the realism of movement without tunneling, qualitative feedback indicated that excessive visual intensity, rather than the absence of tunneling, was the primary cause of discomfort.}

Overall, our study highlights two key findings: (1) the use of locomotion tunneling was generally effective in reducing symptoms for both mild and high susceptibility participants, and (2) participants with high susceptibility demonstrated significant symptom reduction on Day 4 across all tunneling conditions, although the degree of relief varied.

\subsection{Limitations}

\colour{First, although we considered participants’ levels of cybersickness, we did not consider other individual differences such as gender, visual impairments (e.g., myopia, hyperopia), or other perceptual or cognitive factors that may influence cybersickness susceptibility or symptom severity. Future work should incorporate a broader range of participant characteristics to better understand individual variability in adaptation.}


\colour{Second, the experiment involved two different VR headset models (Meta Quest 3 and Meta Quest 3s). Although resolution and performance settings were standardized across devices, and each participant used the same headset throughout the study, hardware-induced variability, such as differences in weight or ergonomics, remains a potential confounding factor.}

\colour{Finally, the VR content used was limited to high motion, simulation style applications, which may not reflect the full spectrum of VR experiences. Future research should examine how various genres, such as narrative driven, exploratory, or meditative VR differ in their impact on cybersickness and mitigation efficacy.}


\section{Conclusion}

This paper investigates whether participants can alleviate cybersickness within a five-day period. We employed VRChat’s VR C1 Runner and VR Jetski Rush as experimental platforms, implementing smooth locomotion and locomotion tunneling for participants with mild and high cybersickness. We employed a between-subjects design and recruited 24 participants. We categorized participants based on their cybersickness severity and recorded their cybersickness feedback over the five days. We found that participants with high cybersickness showed greater symptom alleviation, particularly on Day 4. However, a change in the scene on the fifth day resulted in increased cybersickness among some participants. Furthermore, participants provided qualitative feedback on the causes of cybersickness and the conditions for overcoming it. Finally, we presented the limitations of our current work and outlined plans for future research.

\bibliographystyle{abbrv-doi}

\bibliography{template}
\end{document}